\documentclass{osa-article}
\usepackage[flushleft]{threeparttable}
\usepackage{makecell,multirow}
  
\journal{osac}

\articletype{Research Article}

\begin{document}
	
	\title{Reconstruction of high-resolution 6$\times$6-mm OCT angiograms using deep learning}
	
	\author{Min Gao,\authormark{1} Yukun Guo,\authormark{1} Tristan T. Hormel ,\authormark{1} Jiande Sun,\authormark{2} Thomas Hwang,\authormark{1} and Yali Jia\authormark{1,3,*}}
	
	\address{\authormark{1}Casey Eye Institute, Oregon Health \& Science University, Portland, OR 97239, USA\\
		\authormark{2}School of Information Science and Engineering, Shandong Normal University, Jinan 250358, China\\
		\authormark{3}Department of Biomedical Engineering, Oregon Health \& Science University, Portland, OR 97239, USA}
	
	\email{\authormark{*}jiaya@ohsu.edu} 
	
	
	
	\begin{abstract}
	Typical optical coherence tomographic angiography (OCTA) acquisition areas on
	commercial devices are 3$\times$3- or 6$\times$6-mm. Compared to 3$\times$3-mm angiograms with proper
	sampling density, 6$\times$6-mm angiograms have significantly lower scan quality, with reduced
	signal-to-noise ratio and worse shadow artifacts due to undersampling. Here, we propose a
	deep-learning-based high-resolution angiogram reconstruction network (HARNet) to generate
	enhanced 6$\times$6-mm superficial vascular complex (SVC) angiograms. The network was trained
	on data from 3$\times$3-mm and 6$\times$6-mm angiograms from the same eyes. The reconstructed 6×6-
	mm angiograms have significantly lower noise intensity, stronger contrast and better vascular
	connectivity than the original images. The algorithm did not generate false flow signal at the
	noise level presented by the original angiograms. The image enhancement produced by our
	algorithm may improve biomarker measurements and qualitative clinical assessment of 6$\times$6-
	mm OCTA.
	\end{abstract}
	
	\section{Introduction}
	Optical coherence tomographic angiography (OCTA) is a non-invasive imaging technology that can capture retinal and choroidal microvasculature $in vivo$ \cite{jia2015quantitative}. Clinicians are rapidly
	adopting OCTA for evaluation of various diseases, including diabetic retinopathy (DR) \cite{hwang2015optical,rosen2019earliest}, age-related macular degeneration (AMD) \cite{jia2014quantitative,roisman2016optical}, glaucoma \cite{takusagawa2017projection,rao2016regional}, and retinal vessel occlusion (RVO) \cite{patel2018plexus,tsuboi2019collateral}.High-resolution and large-field-of-view OCTA improve clinical observations, provide useful biomarkers and enhance the understanding of retinal and choroidal microvascular circulations \cite{de2015review,jia2017wide,ishibazawa2019retinal,you2019detection}. Many enhancement techniques have been applied to improve the OCTA image quality, including a regression-based algorithm bulk motion subtraction in OCTA \cite{camino2017regression}, multiple en face image averaging \cite{uji2018multiple,camino2016automated}, enhancement of morphological and vascular features using a modified Bayesian residual transform \cite{tan2018enhancement}, and quality improvement with elliptical directional filtering \cite{chlebiej2019quality}.These approaches can improve vessel continuity and suppress the background noise on angiograms with proper sampling density (i.e., sampling density that meets the Nyquist criterion). However, while commercial systems offer a range of fields of view, only 3$\times$3-mm angiograms are adequately sampled for capillary resolution as the OCTA system scanning speed limits the number of A-lines included on each cross-sectional B-scan. Conventional image enhancement techniques like those mentioned above are not effective on the under-sampled 6$\times$6-mm-mm angiograms. This is unfortunate since the larger scans, with reduced resolution, are in more need of enhancement. The difficulty in enlarging the field without sacrificing resolution is a significant issue for development of OCTA technology, as its field of view is significantly smaller than modalities such as fluorescein angiography (FA).
	
	Recently, deep learning has achieved dramatic breakthroughs, and researchers have proposed a number of convolutional neural networks (CNN) for OCTA image processing \cite{prentavsic2016segmentation,guo2018mednet,nagasato2019automated,guo2019automatic,guo2019development,lauermann2019automated,wang2020automated,wang2020robust}. As an important branch of image processing, super-resolution image reconstruction and enhancement also benefited from deep-learning-based methods \cite{kim2016accurate,ledig2017photo,tong2017image,xu2018dense,zhang2019deep}. Here, we propose a high-resolution angiogram reconstruction network (HARNet) to reconstruct high-resolution angiograms of the superficial vascular complex (SVC). We evaluated the reconstructed high-resolution OCTA for noise level in the foveal avascular zone (FAZ), contrast, vascular connectivity, and false flow signal. We also demonstrate that HARNet is capable of improving not just under-sampled 6$\times$6-mm, but 3$\times$3-mm angiograms as well.
	\section{Methods}
	\subsection{Data acquisition}
	The 6$\times$6- and 3$\times$3-mm OCTA scans of the macula used in this study were acquired with 304$\times$304 A-lines using a 70-kHz commercial OCTA system (RTVue-XR; Optovue, Inc.). Two repeated B-scans were taken at each of the 304 raster positions and each B-scan consisted of 304 A-lines. The split-spectrum amplitude-decorrelation angiography (SSADA) algorithm was used to generate the OCTA data \cite{jia2012split}. The reflectance values on structural OCT and flow values on OCTA were normalized and converted to unitless values in the
	range of [0 255].A guided bidirectional graph search algorithm was employed to segment retinal layer boundaries \cite{guo2018automated} (Fig. 1 A1, B1). 3$\times$3- and 6$\times$6-mm angiograms of the SVC (Fig. 1 A2, B2) were generated by maximum projection of the OCTA signal in a slab including the nerve fiber layer (NFL) and ganglion cell layer (GCL). 
	\begin{figure}[h!]
		\centering\includegraphics[width=14cm]{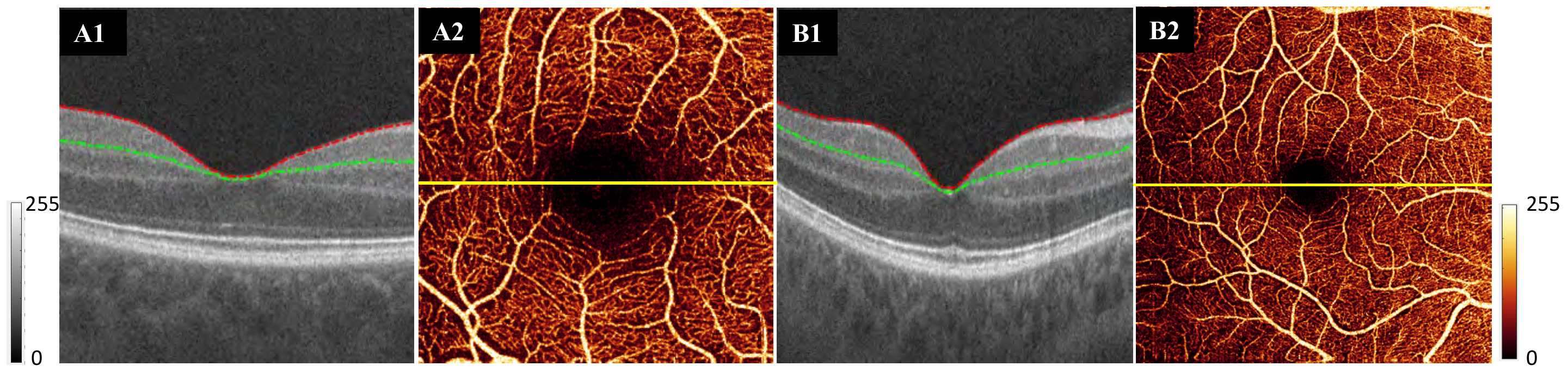}
		\caption{Data acquisition for HARNet. (A1) Cross-sectional structural OCT of a 3$\times$3-mm scan volume, with overlaid boundaries showing the top (red) and bottom (green) of the SVC slab. (A2) 3$\times$3-mm angiogram of the superficial vascular complex (SVC) generated by maximum projection of the OCTA signal in the slab delineated in (A1). The yellow line shows the location of the B-scan in (A1). (B1) and (B2) Equivalent images for 6$\times$6-mm angiograms from the same eye capture more peripheral features, but are of lower quality.}
	\end{figure}
	
	\subsection{Network architecture}
	Our network structure is composed of a low-level feature extraction layer, high-level feature extraction layers, and a residual layer (Fig. 2). Input to the network consists of SVC angiograms. The network first extracts shallow features from the input image through one convolutional layer with 128 channels. Then the high-level features are extracted through four convolutional blocks. Each convolutional block is composed of 20 convolutional layers (C$_{1}$-C$_{20}$) with 64 channels. The kernel size in all the convolutional layers is 3$\times$3 pixels. Skip connections concatenate the output and input of each convolutional block as the input to the next convolutional block. The output and input of the last convolutional block are concatenated and then fed to the residual layer. The residual layer contains a channel that produces the residual image. The residual image and input image are summed to produce the final reconstructed output image. For the most part, low-resolution and high-resolution images have the same low-frequency information, so the output consists of the original input and the residual high-frequency components predicted by HARNet. By only learning these high-frequency components, we were able to improve the convergence rate of HARNet \cite{kim2016accurate}. After each convolutional layer, excluding the residual layer, we added a rectified linear unit (ReLU) \cite{nair2010rectified} to accelerate the convergence of HARNet.
	
	\begin{figure}[h!]
		\centering\includegraphics[width=13cm]{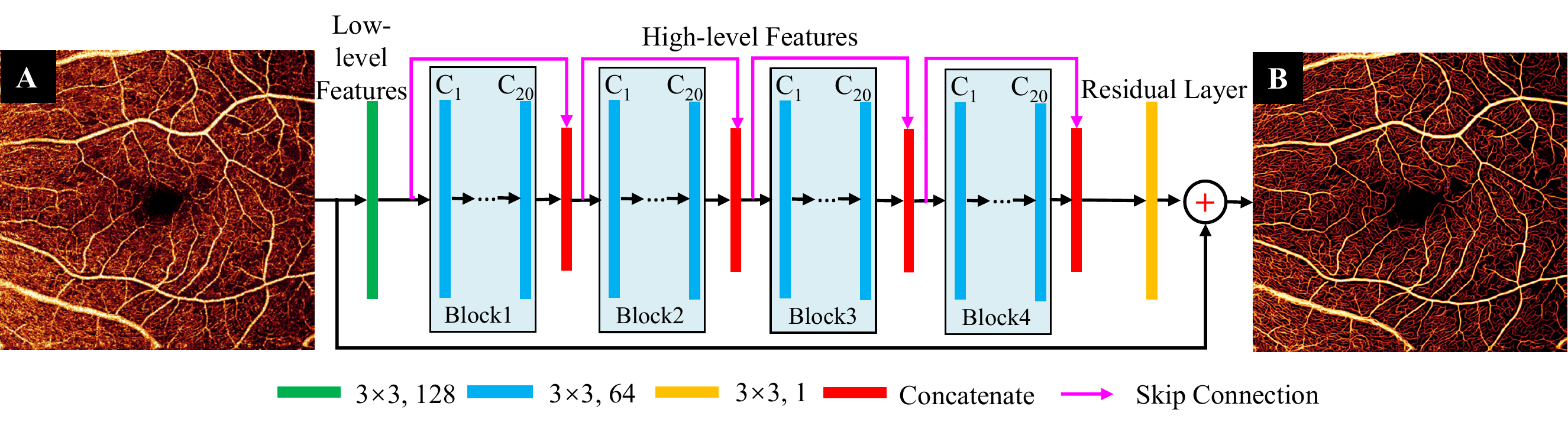}
		\caption{Algorithm flowchart. The network is comprised of three parts: a low-level feature extraction layer, high-level feature extraction layers, and a residual layer. The kernel size in all the convolutional layers is 3$\times$3. The number of channels in the green, blue, and yellow convolutional layer are 128, 64, and 1, respectively. Red layers are concatenation layers that concatenate the output of the convolution block with its input via skip connections. (A) Example input and (B) output 6$\times$6-mm angiogram.}
	\end{figure}
	
	\subsection{Training}
	\subsubsection{Training data preprocessing}
	We trained HARNet by reconstructing 6$\times$6-mm angiograms from their densely-sampled 3$\times$3-mm equivalents. To do so, we first used bi-cubic interpolation to scale the size of the 6$\times$6-mm SVC angiograms (Fig. 3A) by a factor of 2, so that they would be on the same scale as a 3$\times$3-mm scan. Then we used intensity-based automatic image registration \cite{klein2009elastix} (Fig. 3D) to register the scaled 6$\times$6-mm angiograms (Fig. 3B) with the 3$\times$3-mm angiograms (Fig. 3C).The registration algorithm can produce a transform matrix, which contains translation, rotation, and scaling operations. Finally, we cropped the overlapping region from each by taking the maximum inscribed rectangle to construct the input for HARNet and the ground truth (Fig. 3E and F). 
	
	\begin{figure}[h!]
		\centering\includegraphics[width=13cm]{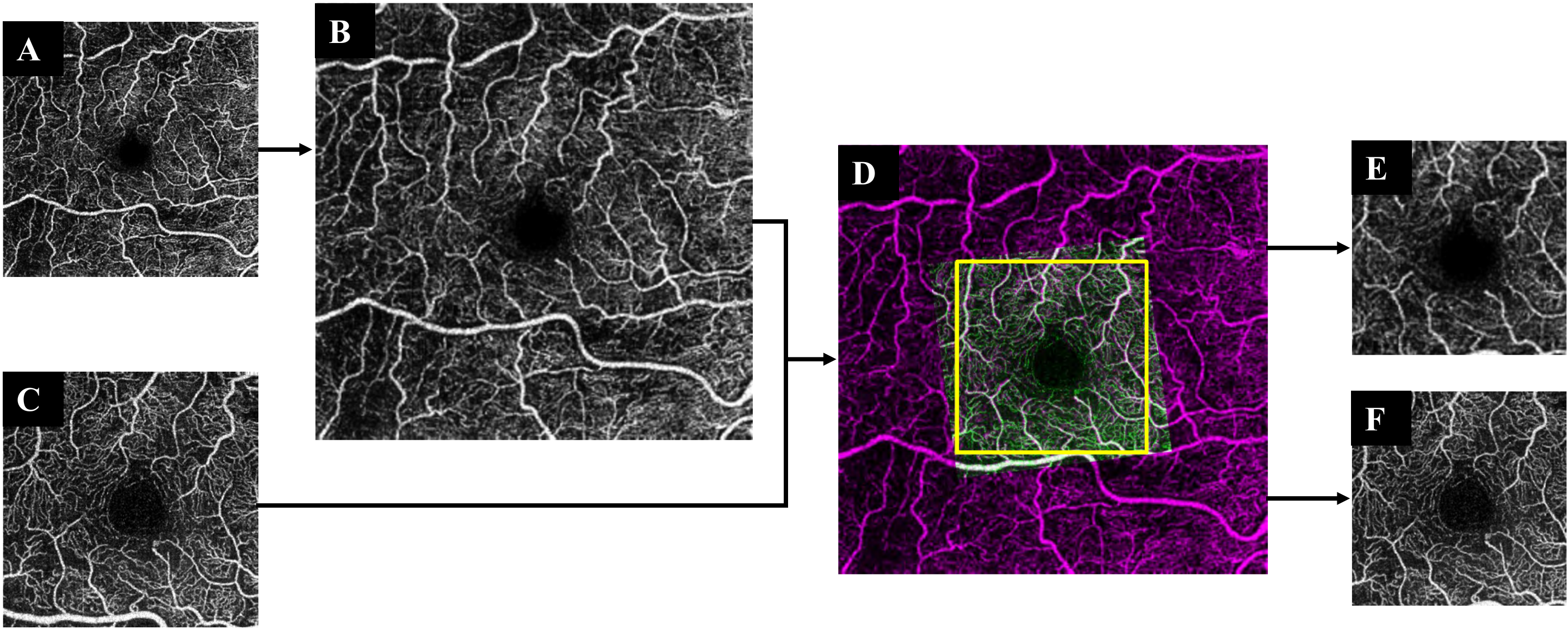}
		\caption{Data preprocessing flow chart. (A) The original 6$\times$6-mm superficial vascular complex (SVC) angiogram. (B) Up-sampled 6$\times$6-mm SVC angiogram. (C) Original 3$\times$3-mm SVC angiogram. (D) Registered image combining both angiograms. The yellow box is the largest inscribed rectangle. (E) Cropped central 3$\times$3-mm section from the 6$\times$6-mm angiogram. (F) The cropped original 3$\times$3-mm angiogram.}
	\end{figure}

	\subsubsection{Loss function}
	We trained the network on a ground truth composed of the original 3$\times$3-mm angiograms filtered with a bilateral filter. To minimize the difference between the output of network and the ground truth, the loss function used in the learning stage was a linear combination of the mean square error (MSE; Eq. 1) and the structural similarity (SSIM; Eq. 2) index \cite{hore2010image,wang2004image}. MSE is used to measure the pixel-wised difference, and SSIM is based on three comparison measurements: reflectance amplitude, contrast, and structure:
	
	\begin{equation}
	\mathrm{MSE}=\frac{1}{w \times h} \sum_{i=1}^{w} \sum_{j=1}^{h}(X(i, j)-Y(i, j))^{2}
	\end{equation}
	
	\begin{equation}
	\mathrm{SSIM}=\frac{2 \mu_{X} \mu_{Y}+C_{1}}{\mu_{X}^{2}+\mu_{Y}^{2}+C_{1}} \cdot \frac{2 \sigma_{X Y}+C_{2}}{\sigma_{X}^{2}+\sigma_{Y}^{2}+C_{2}}
	\end{equation}
	
	\begin{equation}
	\text { Loss }=\mathrm{MSE}+(1-\mathrm{SSIM})
	\end{equation}
	
	where $w, h$ refer to the width and height of the image, $X$ and $Y$ refer to the output of HARNet and the ground truth, respectively, and $\mu_{X}$ and $\mu_{Y}$  are their mean pixel values, $\sigma_{X}$  and $\sigma_{Y}$  are their standard deviations, and $\sigma_{X Y}$ is the covariance. The values of the constants $C_{1}=0.01$ and $C_{2}=0.03$ were taken from the literature \cite{wang2004image}. The loss function (Eq. 3) was a linear combination of the MSE and the SSIM. 
	
	\subsubsection{Subjects and training parameters }
	The data set used in this study consisted of 298 eyes scanned from 196 participants. Each eye was scanned with both a 3$\times$3-mm and a 6$\times$6-mm scan pattern. Ten healthy eyes from 10 participants were intentionally defocused and used in defocusing experiments. Of the remaining 288, we used 210 of these paired scans (randomly selected) for training, and reserved the rest for testing (N=78). The training data includes eyes with DR (N=195) and healthy eyes (N=15). The performance of this network on testing data was separately evaluated on eyes with diabetic retinopathy (N=53) and healthy controls (N=25). Finally, false-flow generation experiments also used 10 cases from the test set of healthy eyes. We used several data augmentation methods to expand the training dataset, including horizontal flipping, vertical flipping, transposition, and 90-degree rotation. For training, considering the hardware capability and computation cost, we used 38×38-pixel sub-images. To avoid the gradient exploding problem, we normalized the pixel value range to 0-1 using Eq. 4,
	\begin{equation}
	S^{\prime}(i, j)=\frac{S(i, j)-\min (S)}{\max (S)-\min (S)}
	\end{equation}
	where $S(i, j)$ is the pixel value ranging from 0-255 at position $(i, j)$ of the angiogram, $S^{\prime}(i, j)$ is the normalized pixel value at location $(i, j)$, and $\min (\cdot)$ and $\max (\cdot)$ are minimum and maximum pixel value of overall image, respectively. Thus the 1050-images in the training dataset after augmentation can be decomposed into 174,555 sub-images, which are extracted from cropped SVC angiograms with a stride of 19. Since HARNet is a fully convolutional neural network, it can be applied on images of arbitrary sizes. Thus, we input the entire image to the model for testing, as the entire image is the clinically relevant data.
	
	An Adam optimizer \cite{kingma2014adam} with an initial learning rate of 0.01 was used to train HARNet by minimizing the loss. We used a global learning rate decay strategy to reduce the learning rate during training in which the learning rate was reduced by $90 \%$ when the loss showed no decline after 2 epochs, provided the rate was greater than $1 \times 10^{-6}$. Training ceased when loss didn$'$t change by more than $1 \times 10^{-5}$ in 3 epochs. The training batch size was 128.
	
	We implemented HARNet in Python 3.6 with Keras (Tensorflow-backend) on a PC with a 16G RAM and Intel i7 CPU, and two NVIDIA GeForce GTX 1080Ti graphics cards.
	
	\section{Results}
	To validate the performance of our algorithm, we used a test dataset that composed of 78 paired original 3$\times$3- and 6$\times$6-mm angiograms and evaluated the reconstructed 3$\times$3-mm and 6$\times$6-mm angiograms using three metrics: noise intensity in the FAZ, global contrast, and vascular connectivity. In addition, we also performed experiments on defocused SVC angiograms, angiograms with different simulated noise intensities, and DR angiograms. 
	
	\subsection{Evaluation metrics}
	\subsubsection{Noise intensity}
	In healthy eyes, the FAZ is avascular, so to obtain an estimate of noise intensity $I_{\text {Noise }}$, we consider the pixel values in 0.3-mm diameter circle R centered in the FAZ
	\begin{equation}
	I_{\text {Noise }}=\frac{1}{R} \times \sum_{(i, j) \in R} S(i, j)^{2}
	\end{equation}
	
	where $S(i, j)$ is the pixel value at position $(i, j)$. 
	
	\subsubsection{Contrast}
	The global contrast of the SVC angiograms produced by the network was measured by the root-mean-square (RMS) contrast \cite{peli1990contrast},
	
	\begin{equation}
	C_{\mathrm{RMS}}=\sqrt{\frac{1}{A} \times \sum_{(i, j) \in A}[S(i, j)-\mu]^{2}}
	\end{equation}
	
	where $S(i, j)$ is the pixel value at position $(i, j)$. $A$ is the total area of the SVC angiogram and $\mu$ is its its mean value.
	
	\subsubsection{Vascular connectivity}
	We also assessed vascular connectivity. To do so, we first binarized the angiograms (Fig. 4 A2-D2) using a global adaptive threshold method \cite{otsu1979threshold}, then skeletonized the binary map to get the vessel skeleton map (Fig. 4 A3-D3). Connected flow pixels were defined as any contiguous flow region with at a length of at least 5 (including diagonal connections), and the vascular connectivity was defined as the ratio of the number of connected flow pixels to the total number of pixels on the skeleton map \cite{jia2012split}. 
	\begin{figure}[h!]
		\centering\includegraphics[width=12cm]{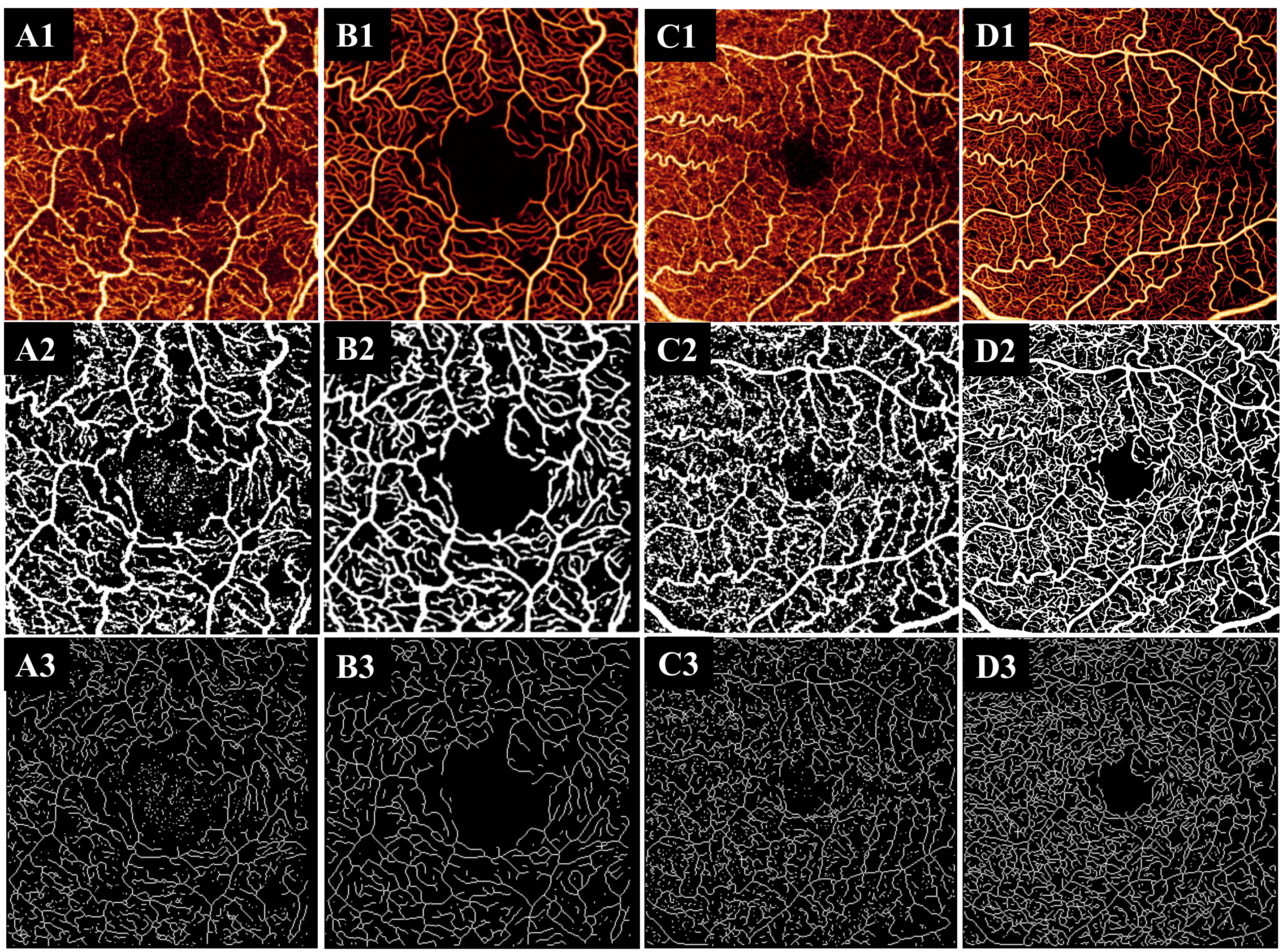}
		\caption{The performance of HARNet. Row 1: (A1) Original 3$\times$3-mm superficial vascular complex (SVC) angiogram and (B1) HARNet output from (A1). (C1) Original 6$\times$6-mm angiogram, and (D1) HARNet output from (C1). Row 2: adaptive threshold binarization of the corresponding images in row 1. Row 3: skeletonization of the corresponding images in row 2. HARNet outputs show enhanced connectivity relative to the original images.}
	\end{figure}

	\subsection{Performance on defocused angiograms}
	
	In order to further verify that our algorithm can improve the image quality of low-quality scans, we also evaluated its performance on defocused angiograms. To obtain defocused scans, we first performed autofocus to optimize the focal length to get optimal scans, and then manually adjusted the focal length to obtain angiograms defocused by 3 diopters. Finally, 10 defocused 3$\times$3-mm angiograms and 12 defocused 6$\times$6-mm angiograms were obtained. Defocused angiograms have lower signal-to-noise ratios than correctly focused angiograms, and vessels also appear dilated. The results show that angiograms reconstructed from defocused 3$\times$3- and 6$\times$6-mm angiograms had lower noise intensity and better connectivity than scans acquired under optimal focusing conditions (Fig. 5, Table 1). Therefore, our algorithm is also applicable to defocused angiograms and improves the quality of such scans. Since defocus leads to a general reduction in scan quality, this result also implies that our algorithm could be used to clean low-quality scans.
	
	\begin{figure}[h!]
		\centering\includegraphics[width=8cm,height=8cm]{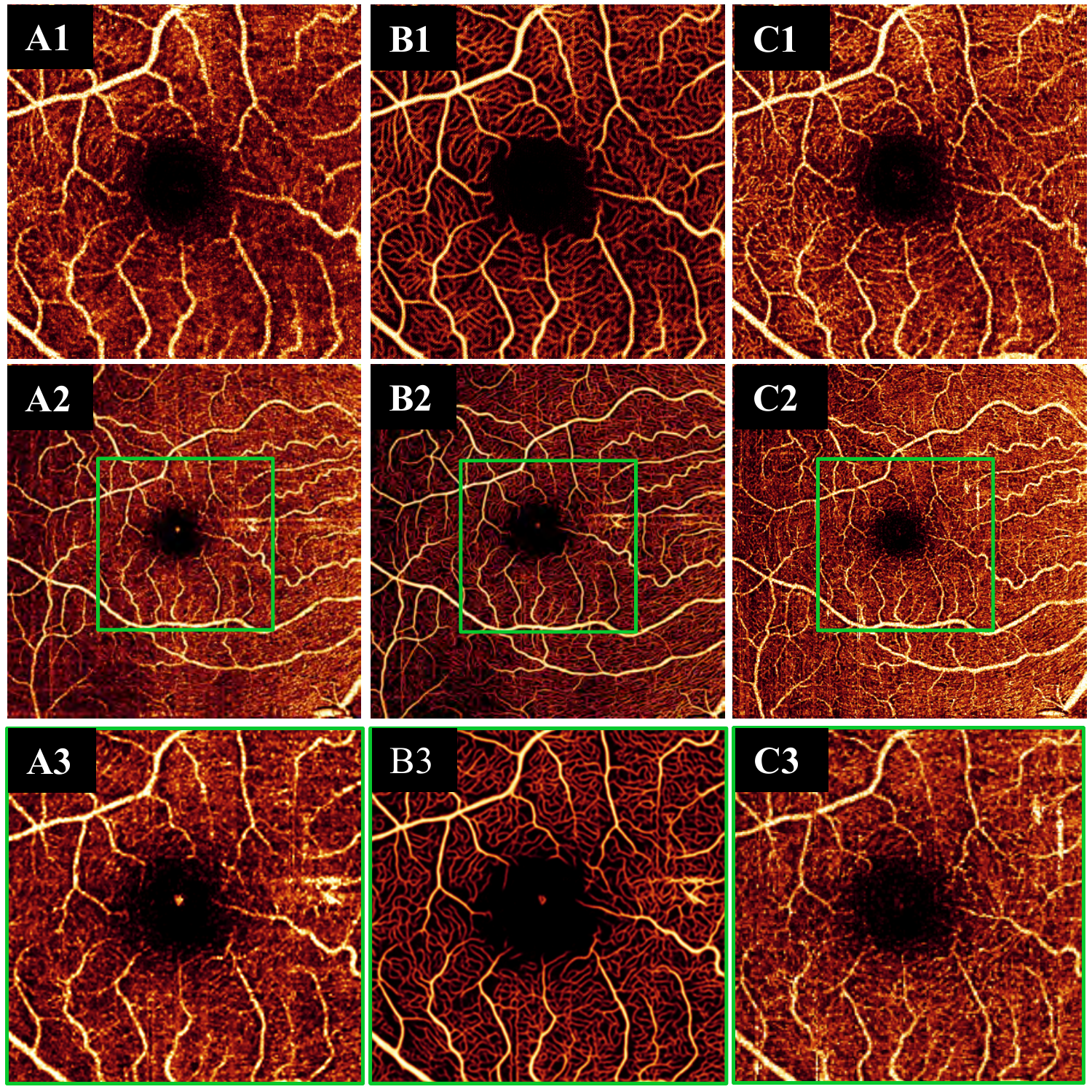}
		\caption{Qualitative demonstration of image quality improvement by the proposed reconstruction method. (A1) 3 diopter defocused 3$\times$3-mm superficial vascular complex (SVC) angiogram. (B1) Reconstruction of (A1). (C1) 3$\times$3-mm OCTA acquired under optimal conditions. (A2) 3 diopter defocused 6$\times$6-mm SVC angiogram. (B2) Reconstruction of (A2). (C2) 6$\times$6-mm angiogram acquired under optimal conditions. (A3) Central 3$\times$3-mm section from the defocused 6$\times$6-mm SVC angiogram. (B3) Reconstruction of (A3). (C3) Central 3$\times$3-mm section from the 6$\times$6-mm angiogram acquired under optimal focusing conditions. The green box is the central 3$\times$3-mm section in the 6$\times$6-mm SVC angiograms.}
	\end{figure}

	\begin{table}
		\centering
		\footnotesize
		\captionsetup{width=13cm}
		\caption{Noise intensity, contrast and vascular connectivity (mean $\pm$ std.) of reconstructed defocused SVC angiograms, and angiograms captured under optimal conditions.}
		\begin{threeparttable}
			\begin{tabular}{ p{1cm} p{4.5cm} p{3cm} p{1.75cm} p{1.75cm} } 
				\hline & &\textbf{ \makecell{Noise intensity\\}} & \textbf{\makecell{Contrast \\}} & \textbf{\makecell{Connectivity\\}} \\
				\hline \multirow{4.5}{4em}{\textbf{\makecell{3$\times$3-mm\\(N$=$10)}}}& \textbf{\makecell{SVC angiograms with optimal focuses}} &\makecell{ $80.89 \pm 79.87$} & \makecell{$54.70 \pm 1.29$} &\makecell{ $0.83 \pm 0.04$} \\
				&\textbf{\makecell{Reconstructed SVC on defocused angiograms}} & \makecell{$1.12 \pm 0.56$} & \makecell{$56.33 \pm 2.23$} & \makecell{$0.91 \pm 0.02$} \\
				& \textbf{\makecell {Optimal vs. Reconstructed}} &\makecell {P<0.001 \\(Mann-Whitney $U$ test\tnote{*})} & \makecell{$\mathrm{P}=0.216$ \\(t-test)} & \makecell{$\mathrm{P}<0.001$ \\(t-test)}\\
				
				\hline \multirow{4.5}{4em}{\textbf{\makecell{6$\times$6-mm\\(N$=$10)}}}&\textbf{\makecell{ SVC angiograms with optimal focuses}} & \makecell{$139.71 \pm 86.90$} &\makecell{ $56.64 \pm 2.00$} & \makecell{$0.78 \pm 0.02$} \\
				& \textbf{\makecell{Reconstructed SVC on defocused angiograms}}& \makecell{$1.28 \pm 3.50$} & \makecell{$56.00 \pm 2.17$} &\makecell{ $0.95 \pm 0.01$ }\\
				& \textbf{\makecell{Optimal vs. Reconstructed}}& \makecell {P<0.001\\(Mann-Whitney $U$ test\tnote{*})}& \makecell {$\mathrm{P}=0.856$\\(t-test) } & \makecell {P<0.001\\(t-test)}  \\
				\hline
			\end{tabular}
			\begin{tablenotes}
				\item[*]The Shapiro-Wilk test was used to check for normality of all variables. Mann-Whitney U test was used for noise intensity that deviates statistically significantly from a normal distribution. N is the number of eyes.
			\end{tablenotes}
		\end{threeparttable}
	\end{table}

	\subsection{Assessment of False flow signal }
	One concern in OCTA reconstruction is the generation of false flow signal. Because OCTA reconstruction methods are designed to enhance vascular detail, they are susceptible to mistakenly enhancing background that may randomly share some features with true vessels. In order to evaluate whether HARNet produces such artifacts, we selected 10 3$\times$3-mm angiograms with good quality from 10 healthy eyes and then produced denoised angiograms by applying a simple Gabor and median filter to the original 3$\times$3-mm angiograms (Fig. 6 A1). Then we added Gaussian noise to the denoised angiograms using different parameters $(\mu, \sigma)$ (Fig .6 B1-E1). We varied $\mu$ and $\sigma$ separately in increments of 0.005 from 0.001 to 0.1 and from 0.001 to 0.05, respectively, to obtain 2000 noisy 3$\times$3-mm SVC angiograms with different noise intensities (0 - 2100). Next, we input the denoised and noisy angiograms into the network to obtain reconstructed angiograms from each (Fig. 6 A2-E2). The false flow signal intensity was defined as 
	
	\begin{equation}
	I_{\text {flase flow signal }}=\frac{1}{R} \times \sum_{(i, j) \in R} S(i, j)^{2}
	\end{equation}
	
	where $I_{\text {flow signal }}$ is the false flow signal intensity, and R corresponds to the same, physiologically flow-free 0.3-mm diameter circle within the FAZ as previously. We found our algorithm did not generate false flow signal when the noise intensity was under 500, which is far above the noise intensity measured in original 3$\times$3-mm (146.77 $\pm$ 145.87) and 6$\times$6-mm (93.10 $\pm$ 159.05) angiograms (Fig .7).
	
	\begin{figure}[h!]
		\centering\includegraphics[width=13cm]{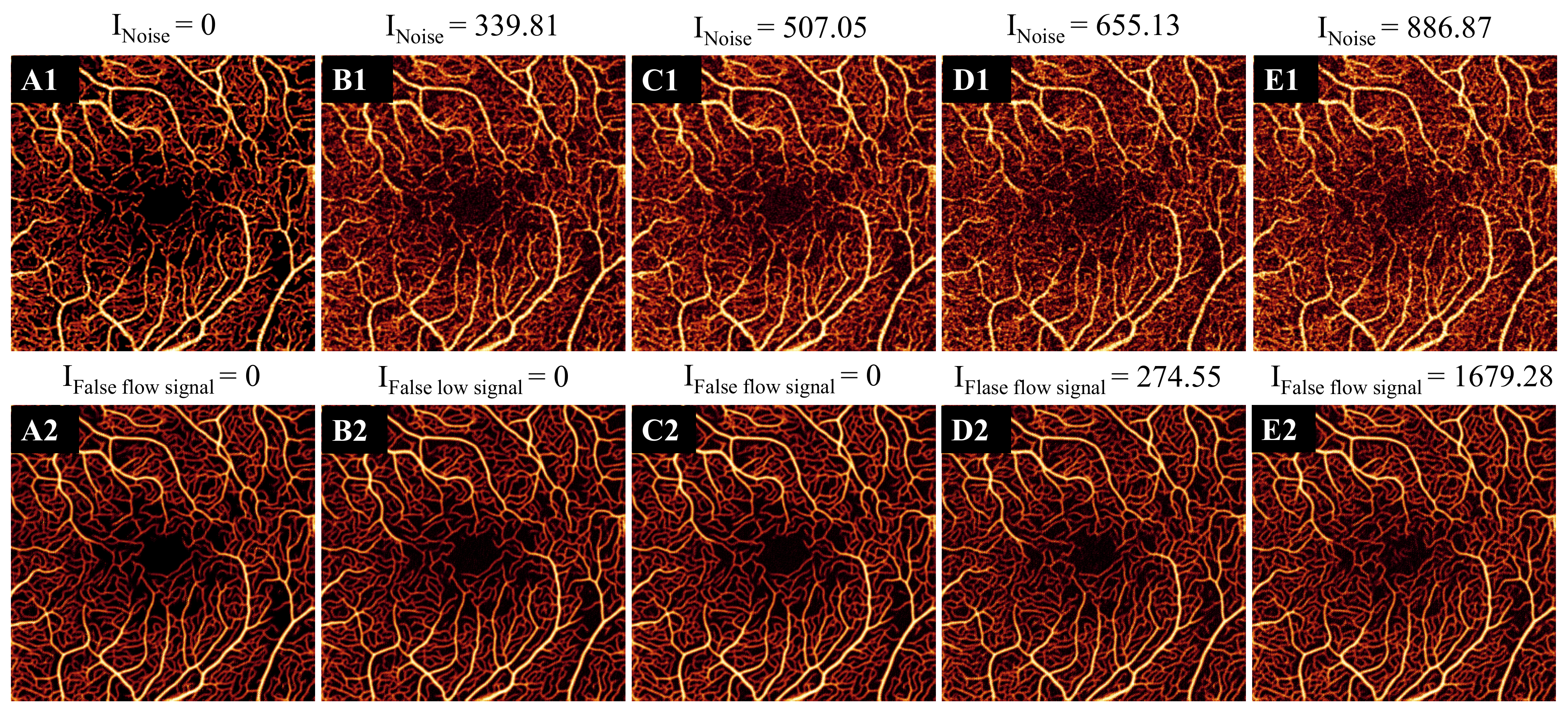}
		\caption{3$\times$3-mm superficial vascular complex (SVC) angiograms with different noise intensities. (A1) In 3$\times$3-mm angiograms denoised with Gabor filtering, the noise intensity is 0. (B1-E1) 3$\times$3-mm SVC angiograms with different noise intensities. (A2-E2) 3$\times$3-mm angiograms reconstructed from the corresponding angiograms in row 1. When the noise intensity is less than 500, there is no false flow signal. }
	\end{figure}
	
	\begin{figure}[h!]
		\centering\includegraphics[width=12cm]{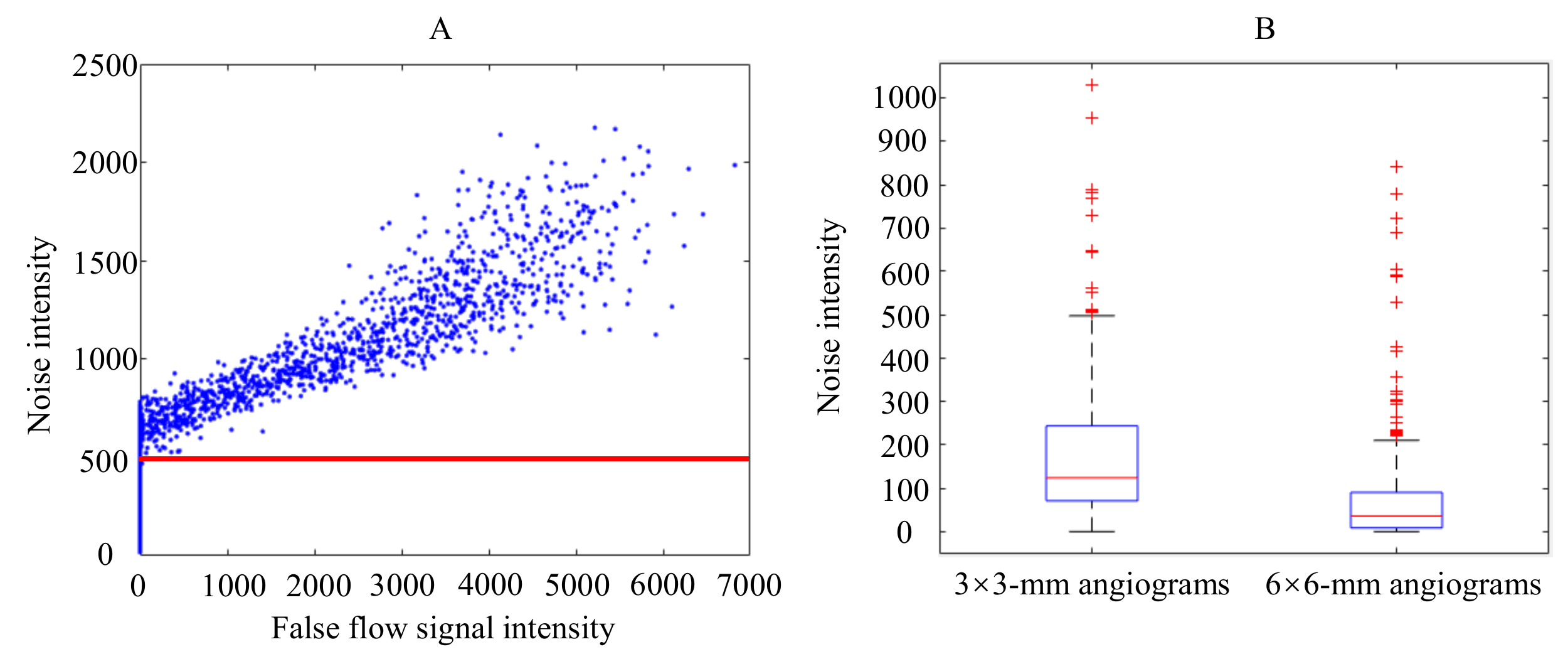}
		\caption{3$\times$3-mm superficial vascular complex (SVC) angiograms with different noise intensities. (A1) In 3$\times$3-mm angiograms denoised with Gabor filtering, the noise intensity is 0. (B1-E1) 3$\times$3-mm SVC angiograms with different noise intensities. (A2-E2) 3$\times$3-mm angiograms reconstructed from the corresponding angiograms in row 1. When the noise intensity is less than 500, there is no false flow signal. }
	\end{figure}

	\subsection{Performance on DR angiograms}
	
	Many diseases present outside of the central area of the macula. The enhancement of larger field-of-view angiograms resolution and image quality may improve the measurements of disease biomarkers such as non-perfusion area and vessel density, thereby further helping ophthalmologists diagnose such diseases. However, since features in diseased eyes may vary from healthy, it is possible that image reconstruction algorithms could suffer from reduced performance on such images. To investigate, we examined reconstructed 6$\times$6-mm angiograms (Fig. 8) of eyes with DR, a leading cause of blindness \cite{ogurtsova2017idf}. Although the 6$\times$6-mm angiograms of eyes with DR have higher noise intensity than healthy eyes, results show that the reconstructed DR angiograms also demonstrate the improvement on noise intensity, contrast, and connectivity comparable to that of healthy controls (Table 2). Because abnormal morphological vessels play a very important role in the diagnosis, it is essential to retain the abnormal vascular morphology when processing images. The DR angiograms reconstructed by our algorithm can preserve pathological vascular abnormalities such as intraretinal microvascular abnormalities (IRMA), early neovascularization and microaneurysms [Fig. 8(A2)].
	
	\begin{figure}[h!]
		\centering\includegraphics[width=13cm]{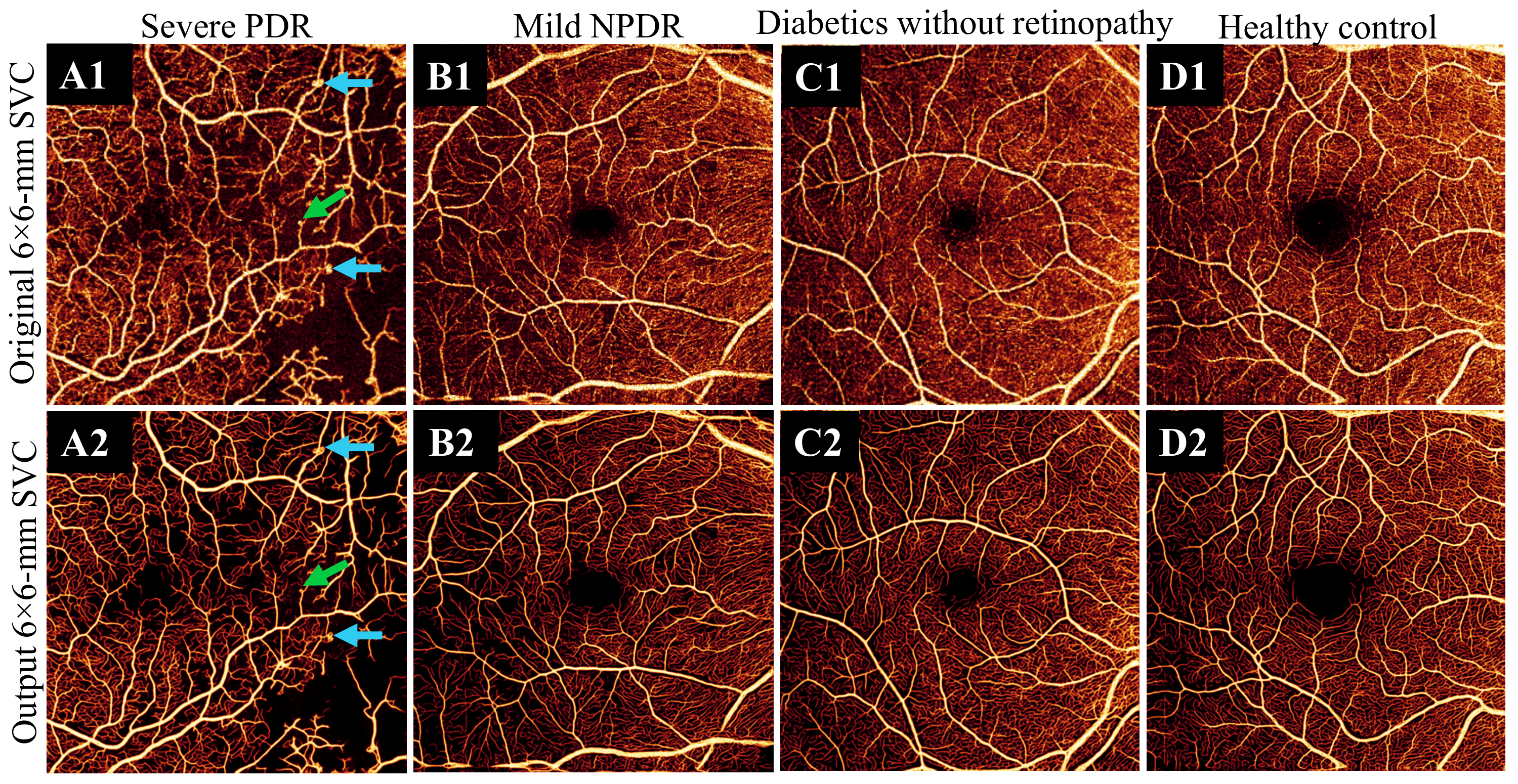}
		\caption{HARNet performance on eyes with DR. Top row: original 6$\times$6-mm superficial vascular complex (SVC) angiograms from an eye with active proliferative diabetic retinopathy (PDR) (A1), severe non-proliferative diabetic retinopathy (NPDR) (B1), diabetics without retinopathy (C1), and a healthy control (D1). Bottom row: 6$\times$6-mm angiograms (A2-D2) HARNet output for (A1-D1).}
	\end{figure}
	
	\begin{table}
		\centering
		\footnotesize
		\captionsetup{width=12cm}
		\caption{Noise intensity, contrast, and vascular connectivity (mean $\pm$ std.) of reconstructed 6×6-mm SVC angiograms in eyes with diabetic retinopathy and healthy controls.}
		\begin{threeparttable}
			
			\begin{tabular}{p{2cm}p{3cm}p{2cm}p{2cm}p{2cm} }
				\hline & & \textbf{\makecell{Noise intensity\\}} & \textbf{\makecell{Contrast \\}} & \textbf{ \makecell{Connectivity \\}} \\
				\hline \multirow{3}{4em}{\textbf{\makecell{Healthy controls\\(N$=$25)}}}& \textbf{\makecell{Original}} & \makecell{$50.77 \pm 59.39$}&\makecell{$55.61 \pm 1.23$}&\makecell{$0.80 \pm 0.02$}\\
				& \textbf{\makecell{Reconstructed}} & \makecell{$0.16 \pm 0.26$} &\makecell{$59.61 \pm 1.71$}& \makecell{ $0.96 \pm 0.01$}\\
				&\textbf{\makecell{Improvement}} &\makecell{ $99.69 \%$} & \makecell{$7.20 \%$ }&\makecell{$19.51 \%$}\\
				
				\hline \multirow{3}{4em}{\textbf{\makecell{Diabetic retinopathy\\ (N$=$53)}}} & \textbf{\makecell{Original}} &\makecell{ $109.63 \pm 106.34$} & \makecell{$55.44 \pm 2.65$}&\makecell{$0.80 \pm 0.02$}\\
				& \textbf{\makecell{Reconstructed}} & \makecell{$9.76 \pm 26.76$} &\makecell{ $58.11 \pm 2.68$} & \makecell{$0.97 \pm 0.01$} \\
				& \textbf{\makecell{Improvement}} & \makecell{$91.10 \%$} & \makecell{$8.57 \%$} &\makecell{ $21.06 \%$} \\
				\hline
			\end{tabular}
			
		\end{threeparttable}
	\end{table}

	\subsection{Performance of different methods}

	We also compared our algorithm with commonly used image enhancement methods including Gabor and Frangi filters. Compared to the original angiograms, our method significantly reduces noise and improves the vascular connectivity without producing false flow signal on all sizes of scans. There is no significant improvement of image contrast on 3×3-mm scans [(Fig. 9(D1)], while, the contrast shows significant improvement on 6×6-mm scans [Fig. 9(D2); Table 3]. The Gabor filter reduces the noise intensity and improves vascular connectivity, but the contrast is greatly reduced [Figs. 9(B1), 9(B2); Table 3]. The Frangi filter significantly enhances the contrast and improves vascular connectivity, but the noise intensity is significantly increased and may produce false flow signal [Figs. 9(C1), 9(C2); Table 3]
	\begin{figure}[h!]
		\centering\includegraphics[width=13cm]{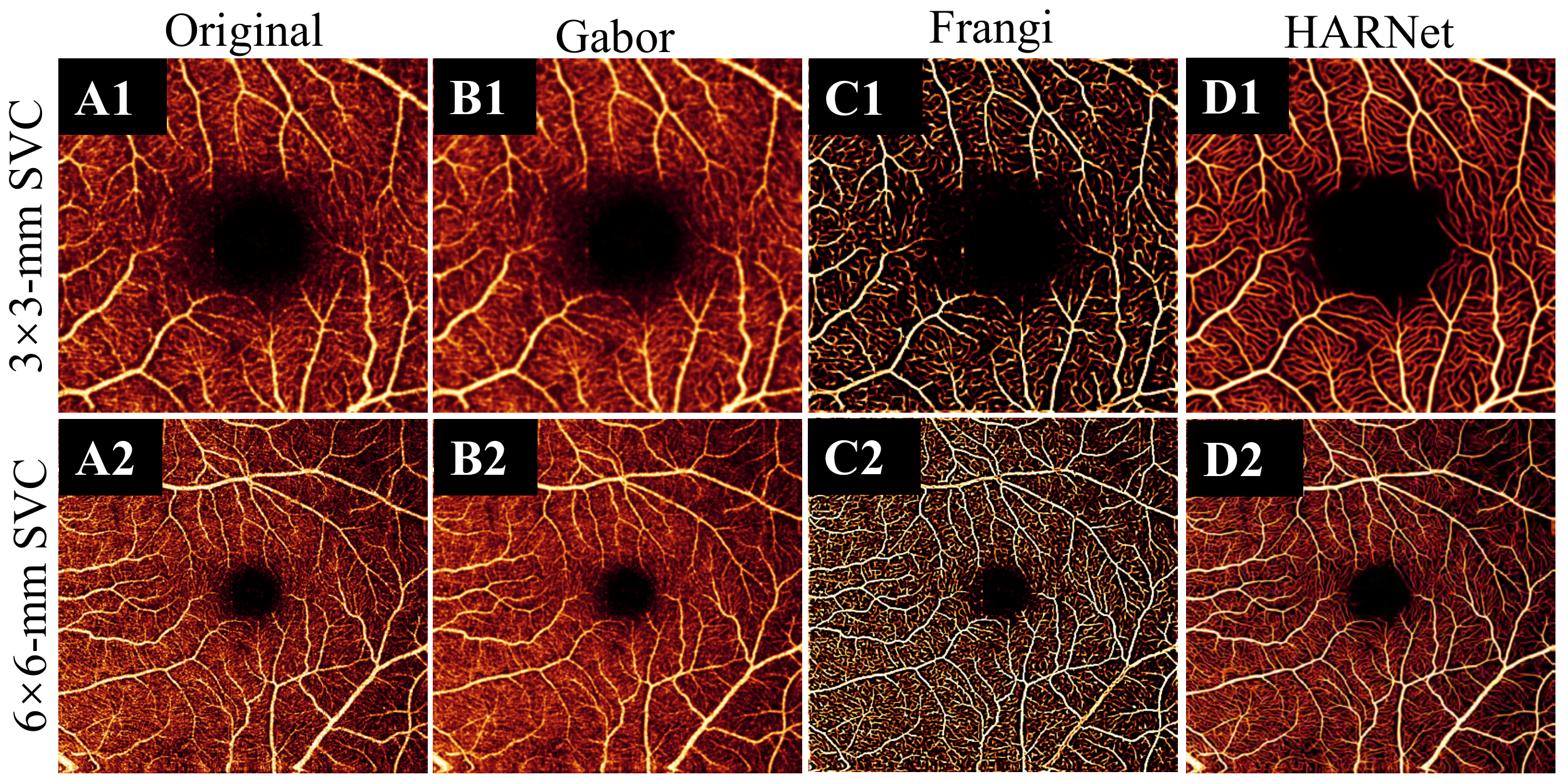}
		\caption{Performance of different methods on image enhancement. Top row: original 3×3- mm superficial vascular complex (SVC) angiograms from a healthy eye; (A1) original data, (B1) after applying a Gabor filter, (C1) after applying a Frangi filter, and (D1) reconstructed using the proposed method. Bottom row: equivalent from a 6×6-mm scan.}
	\end{figure}
	
	\begin{table}
	\centering
	\footnotesize
	\captionsetup{width=12cm}
	\caption{Comparison of noise intensity, contrast, and vascular connectivity (mean $\pm$ std.) between original angiograms and angiograms processed by different methods. N is the number of eyes.}
		\begin{threeparttable}
			\begin{tabular}{ p{2cm} p{2cm} p{2.75cm} p{2cm} p{2cm} }
				\hline & &\textbf{ \makecell{Noise intensity\\}} & \textbf{\makecell{Contrast \\}} & \textbf{\makecell{Connectivity\\}} \\
				
				\hline \multirow{3.5}{4em}{\textbf{\makecell{3$\times$3-mm\\(N$=$78)}}} &\textbf{Original}& \makecell{$163.87 \pm 136.13$} & \makecell{$53.53 \pm 2.31$} & \makecell{$0.87 \pm 0.03$} \\

				& \textbf{Gabor} & \makecell{$108.15 \pm 94.13$\tnote{*}} & \makecell{$44.97 \pm 2.27$} & \makecell{$0.87 \pm 0.03$\tnote{*}} \\
				
				& \textbf{Frangi} & \makecell{$260.92 \pm 278.00$} & \makecell{$82.42 \pm 4.37$\tnote{*}} & \makecell{$0.90 \pm 0.03$\tnote{*}} \\
				
				& \textbf{HARNet(proposed)} & \makecell{$5.79 \pm 7.83$\tnote{*}} & \makecell{$53.82 \pm 3.05$} & \makecell{$0.93 \pm 0.02$\tnote{*}} \\

				\hline \multirow{3.5}{4em}{\textbf{\makecell{6$\times$6-mm\\(N$=$78)}}} & \textbf{Original} & \makecell{$98.10 \pm 113.12$} & \makecell{$54.19 \pm 2.10$} & \makecell{$0.80 \pm 0.02$} \\
				
				& \textbf{Gabor} & \makecell{$63.92 \pm 83.35$} & \makecell{$44.39 \pm 2.04$} & \makecell{$0.83 \pm 0.02$\tnote{*}} \\
				
				& \textbf{Frangi} & \makecell{$167.17 \pm 268.57$} & \makecell{$82.03 \pm 2.96$\tnote{*}} & \makecell{$0.88 \pm 0.02$\tnote{*}} \\
				
				& \textbf{HARNet(proposed)} & \makecell{$8.22 \pm 24.32$\tnote{*}} & \makecell{$58.59 \pm 2.50$\tnote{*}} & \makecell{$0.96 \pm 0.00$\tnote{*}} \\	
						
				\hline
			\end{tabular}
			\begin{tablenotes}
				\item[*] Compared to original images using paired t test, the validation metrics with significant improvement (P-value<0.001) was annotated with.
			\end{tablenotes}
		\end{threeparttable}
	\end{table}

	\section{Discussion}
	Image analysis of low-quality or under-sampled OCTA is challenging in several respects. Noise affects the visibility of small blood vessels, especially capillaries, leading to artifactual vessel fragmentation. Motion and shadow artifacts are common, and amplified by under-sampling. OCTA quality, then, can have a significant impact on the judgment of ophthalmologists or researchers. To help mitigate this concern, several noise reduction and image enhancement procedures have been proposed. To reduce noise and enhance vascular connectivity, datasets are sometimes obtained by acquiring multiple images of the same location over time, making it possible to apply various averaging techniques \cite{uji2018multiple,camino2016automated,mo2017visualization,maloca2017enhanced}. However, the acquisition of larger and larger amounts of data makes the total acquisition time longer, increasing the probability of image artifacts caused by eye motions and introducing additional difficulty for clinical imaging. Filtering is also often applied to OCTA images to improve image quality \cite{chlebiej2019quality,hendargo2013automated}, but typical problems in data filtering are reduced image resolution and the loss of capillary signal. Other noise reduction strategies suffer similar issues. For instance, a regression-based algorithm [14] that can remove decorrelation noise due to bulk motion in OCTA has been reported. Although image contrast was improved by this method, the drawback is worse vessel continuity, and it also suffers the loss of capillaries with weak signal. 
	
	In this study, our proposed method can not only reduce noise and enhance connectivity, but also improve the capability to resolve capillaries in large-field-of-view scans. The two most common scan patterns used in research and the clinic are 3$\times$3-mm and 6$\times$6-mm \cite{kashani2017optical,spaide2018optical}. While the smaller 3$\times$3-mm OCTA can obtain higher image quality due to the denser scanning pattern, its small fields-of-view is a major limitation. Our algorithm$'$s ability to enhance 6$\times$6-mm OCTA is a step toward compensating for this limitation. We achieved this enhancement by training a network to reconstruct images by learning features from the high-definition 3$\times$3-mm images. This means that we did not need to manually segment vasculature to generate the ground truth, or generate high-definition scans by using a new scanning protocol in a prototype \cite{prentavsic2016segmentation}. Therefore, our approach is a practical method to enhance 6$\times$6-mm images by using an acquired 3$\times$3-mm image, that could in principle also be extended to even larger fields-of-view with sparser sampling. Such enhancement via intelligent software could prove to be a superior method for achieving high-quality, large-field scans since hardware solutions (like, for example, increasing sampling density or incorporating adaptive optics) quickly lead to prohibitive cost and imaging times. Improving image quality and resolution may in turn promote better measurements of disease biomarkers such as non-perfusion area and vessel density; by extending improved image quality to a larger field-of-view we also increase the chance that we will detect pathology since disease can manifest outside of the central macular region usually imaged with OCTA \cite{you2019detection,russell2019distribution}. 
	
	We investigated the quality of our algorithm$'$s output by evaluating reconstructed angiograms with three metrics: noise intensity in the FAZ, global contrast, and vessel connectivity. The angiograms obtained by our algorithm have almost no noise in the FAZ (0.16 $\pm$ 0.26) and vascular connectivity was likewise increased in the HARNet-processed images. In addition to these quantitative improvements, we consider the HARNet output images to appear qualitatively cleaner than the unprocessed input. We also performed experiments on defocused SVC angiograms, and the results show that the algorithm can improve such scans, which is an indication of robustness and broad utility. To demonstrate that the restored flow signal in the reconstructed angiograms is real, we verified whether a false flow signal is generated by using angiograms with different simulated noise intensities. The results show that our algorithm did not generate false flow signal when the noise intensity was under 500. This value far exceeds the noise intensity in the clinically-realistic OCTA angiograms examined in this study. Because the noise intensity in the FAZ and inter-capillary space is similar, we also think that artifactual vessels should not be generated outside of the FAZ. 
	
	HARNet improved the quality of both 3$\times$3- and 6$\times$6-mm OCTA angiograms according to the metrics examined in this study. Specifically, HARNet enhanced the quality of under-sampled 6$\times$6-mm OCTA, while other enhancement algorithms perform poorly on such scans \cite{li2017adaptive,ting2019artificial}. And it is interesting that, while HARNet was trained to reconstruct high-resolution 6$\times$6-mm angiograms from sparsely sampled scans, the network also improved 3$\times$3-mm images. In particular, the angiograms reconstructed from defocused scans compared favorably to equivalent images acquired at optimal focus for both scanning patterns. This implies that HARNet is effective as a general OCTA image enhancement tool, outside of the specific context of 6$\times$6-mm angiogram reconstruction. Additionally, the image improvement provided by HARNet is more than just cosmetic, as demonstrated by the improvement in vessel connectivity. Although beyond the scope of this study, we speculate that other OCTA metrics (e.g., non-perfusion area or vessel density) may also prove to be more accurately measured on HARNet-reconstructed images.
	
	Deep-learning-based algorithms are “black boxes” compared to the conventional image processing algorithms. Interpretability of deep-learning is an important field of research in machine learning. Matthew et al. \cite{zeiler2014visualizing} tried to understand CNNs using a kernel visualization technique. More recently, researchers proposed many methods to explain how CNNs work \cite{zhang2018interpretable,samek2017explainable,zhang2018visual}. For a specific CNN, we can use kernel visualization techniques or heat maps to understand what features the CNN used to make decisions \cite{zintgraf2017visualizing}. For our future work, we could use the same visualization techniques to understand why HARNet is very effective on reconstructing angiograms, and employ an ablation study to get a deeper understanding of the structure of HARNet. The biggest advantage of deep-learning-based methods is that they have strong generalizability, which means CNNs can make a reliable prediction on unseen data. Furthermore, the transfer-learning technique is used to transfer the knowledge learned from one dataset to a new dataset using a small number of samples. OCTA data form different pathologies of the retina share a similar feature space. Thus, with the innate strong generalizability and transfer-learning technique, our HARNet should be able to handle OCTA data from different pathologies of the retina (i.e., age-related-degeneration and glaucoma).
	
	\begin{figure}[h!]
		\centering\includegraphics[width=11cm]{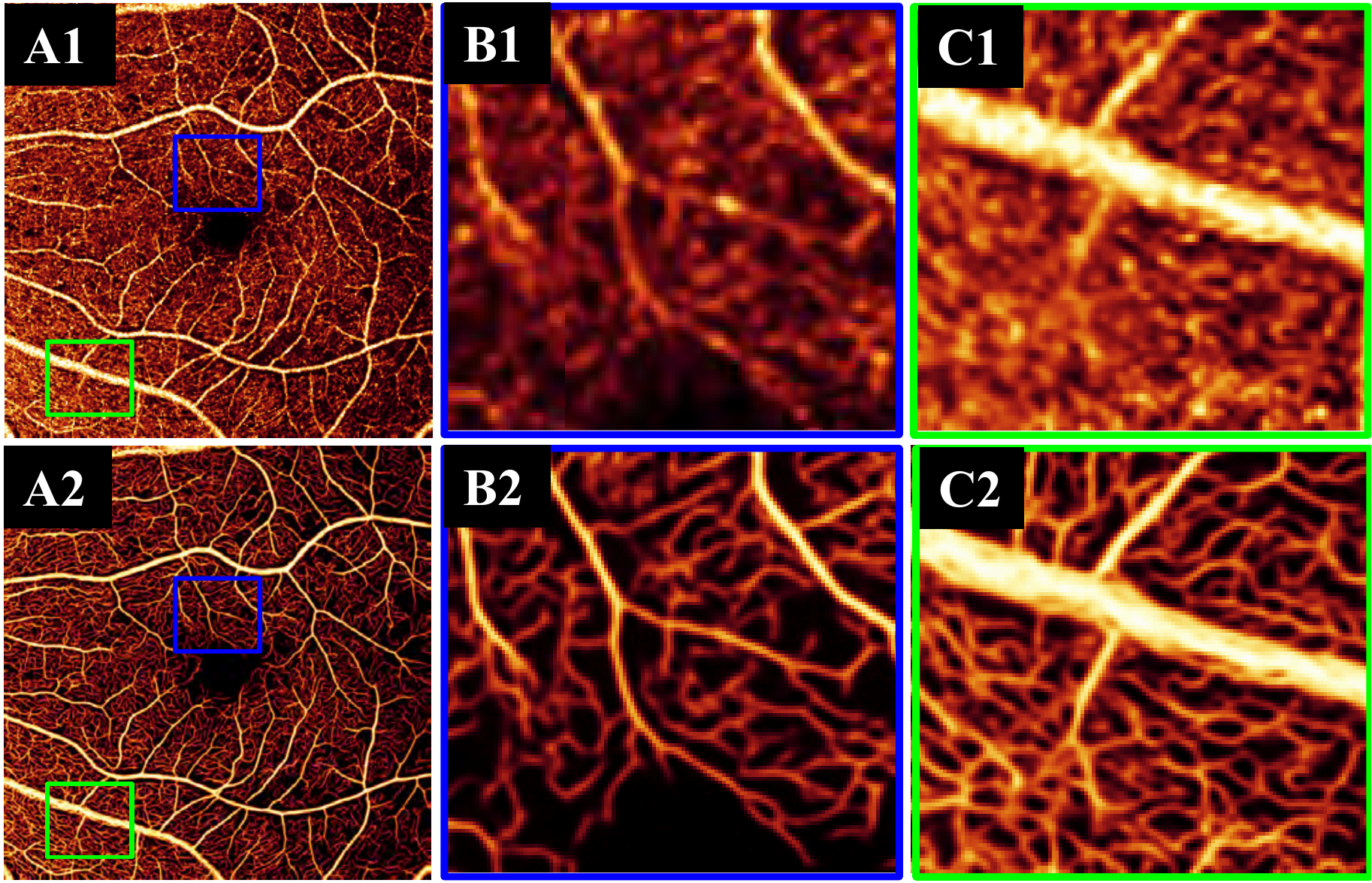}
		\caption{(A1) Original 6$\times$6-mm superficial vascular complex (SVC) angiograms. (B1) The centrally located 3$\times$3-mm angiograms. (C1) The region outside the centrally located 3$\times$3-mm angiograms. (A2-C2) HARNet output for (A1-C1). Since the ground truth used in training did not include the specific vascular patterns present in the green square, the reconstruction here may not be ideal (C2).}
	\end{figure}
	
	There are some limitations to this study. Since we trained HARNet by using optimally sampled, centrally located 3$\times$3-mm angiograms, features specific to the periphery, like for instance the grating-like vascular structure of the radial peripapillary capillaries (Fig. 10 C1), could not be learned during training. HARNet therefore may introduce features that are physiologically specific to the central macula into more peripheral regions (Fig. 10 B2, C2). Likewise, HARNet may remove features specific to peripheral regions, particularly if there are disease-specific features that are more prevalent in the periphery compared to the macula, such as neovascularization elsewhere, which tend to occur more along the major vessels, away from the central macula. Unfortunately, due to the lack of a high-resolution ground truth for the region outside the central macula, we can only speculate on this issue. HARNet also currently only works in only one vascular complex (the superficial), but the intermediate and deep capillary plexus, as well as the choriocapillaris, are important in several diseases \cite{toto2016retinal,chi2017optical,onishi2018importance,hwang2018automated,camino2019detecting,liu2019projection}. Reconstruction of these vascular layers would also be beneficial; however, issues such as shadowing that present preferentially in low-density scanning patterns are only exacerbated in these deeper layers. This makes image reconstruction in these locations significantly more challenging. Finally, to completely characterize HARNet, it will also be important to assess performance on pathological scans. While our data indicates that HARNet can also perform well on DR angiograms, there are of course many other diseases that could be examined for a more thorough assessment. Furthermore, a complete investigation of HARNet’s performance on these diseases would include the extraction of relevant biomarkers to determine if they are more or less accurately measured on reconstructed images. Due to eye motion, OCTA produces bright strip artifacts that are also passed to reconstructed angiograms (Fig. 11). However, our algorithm did not make efforts to correct this disturbance, since commercial systems could remove most motion artifacts by tracking at the scan acquisition level and such artifacts can also be frequently removed by other software means \cite{camino2017regression,zang2016automated}.
	
	\begin{figure}[h!]
	\centering\includegraphics[width=11cm]{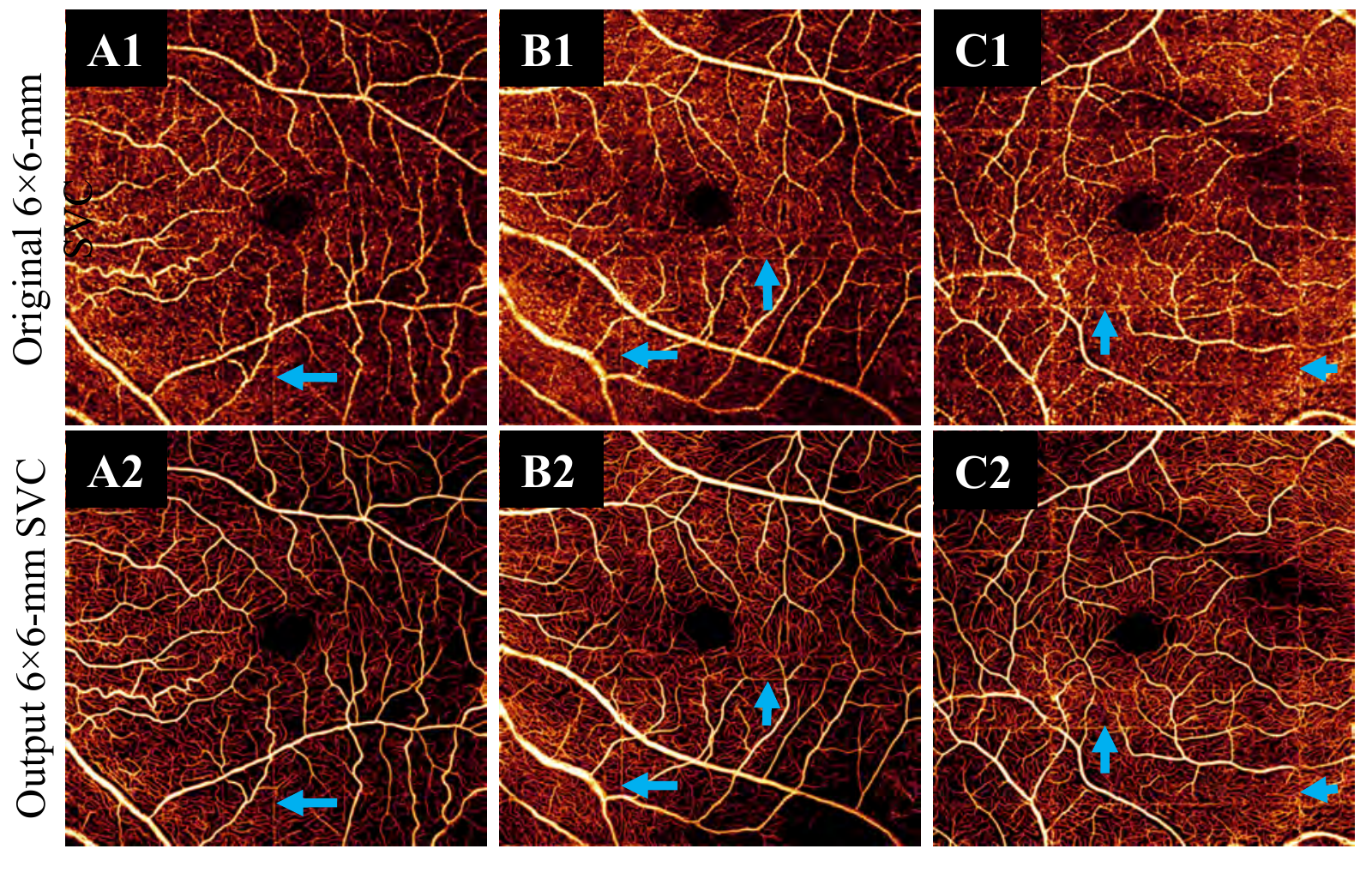}
	\caption{Top row: (A1-C1) Original 6$\times$6-mm superficial vascular complex (SVC) angiograms with motion artifacts. Bottom row: (A2-C2) HARNet output for (A1-C1). Blue arrows indicate the position of motion artifacts.}
	\end{figure}

	\section{Conclusions}
	We proposed an end-to-end image reconstruction technique for high-resolution 6$\times$6-mm SVC angiograms based on high-resolution 3$\times$3-mm angiograms. The high-resolution 6$\times$6-mm angiograms produced by our network had lower noise intensity and better vasculature connectivity than original 6$\times$6-mm SVC angiograms, and we found our algorithm did not generate false flow signal at realistic noise intensities. The enhanced 6$\times$6-mm angiograms may improve the measurements of disease biomarkers such as non-perfusion area and vessel density.
	
	\section*{Funding}
	This work was supported by grants from the National Institutes of Health (R01 EY027833, R01 EY024544, P30 EY010572); an unrestricted departmental funding grant and William \& Mary Greve Special Scholar Award from Research to Prevent Blindness (New York, NY).  
	
	\section*{Disclosures}
	Oregon Health \& Science University (OHSU), Yali Jia has a significant financial interest in Optovue, Inc. These potential conflicts of interest have been reviewed and managed by OHSU.

	\bibliography{manuscript}

\begin{thebibliography}{10}
\newcommand{\enquote}[1]{``#1''}

\bibitem{jia2015quantitative}
Y.~Jia, S.~T. Bailey, T.~S. Hwang, S.~M. McClintic, S.~S. Gao, M.~E. Pennesi,
  C.~J. Flaxel, A.~K. Lauer, D.~J. Wilson, J.~Hornegger \emph{et~al.},
  \enquote{Quantitative optical coherence tomography angiography of vascular
  abnormalities in the living human eye,} {\protect\JournalTitle{Proceedings of
  the National Academy of Sciences}} \textbf{112}, E2395--E2402 (2015).

\bibitem{hwang2015optical}
T.~S. Hwang, Y.~Jia, S.~S. Gao, S.~T. Bailey, A.~K. Lauer, C.~J. Flaxel, D.~J.
  Wilson, and D.~Huang, \enquote{Optical coherence tomography angiography
  features of diabetic retinopathy,} {\protect\JournalTitle{Retina
  (Philadelphia, Pa.)}} \textbf{35}, 2371 (2015).

\bibitem{rosen2019earliest}
R.~B. Rosen, J.~S.~A. Romo, B.~D. Krawitz, S.~Mo, A.~A. Fawzi, R.~E. Linderman,
  J.~Carroll, A.~Pinhas, and T.~Y. Chui, \enquote{Earliest evidence of
  preclinical diabetic retinopathy revealed using optical coherence tomography
  angiography perfused capillary density,} {\protect\JournalTitle{American
  journal of ophthalmology}} \textbf{203}, 103--115 (2019).

\bibitem{jia2014quantitative}
Y.~Jia, S.~T. Bailey, D.~J. Wilson, O.~Tan, M.~L. Klein, C.~J. Flaxel,
  B.~Potsaid, J.~J. Liu, C.~D. Lu, M.~F. Kraus \emph{et~al.},
  \enquote{Quantitative optical coherence tomography angiography of choroidal
  neovascularization in age-related macular degeneration,}
  {\protect\JournalTitle{Ophthalmology}} \textbf{121}, 1435--1444 (2014).

\bibitem{roisman2016optical}
L.~Roisman, Q.~Zhang, R.~K. Wang, G.~Gregori, A.~Zhang, C.-L. Chen, M.~K.
  Durbin, L.~An, P.~F. Stetson, G.~Robbins \emph{et~al.}, \enquote{Optical
  coherence tomography angiography of asymptomatic neovascularization in
  intermediate age-related macular degeneration,}
  {\protect\JournalTitle{Ophthalmology}} \textbf{123}, 1309--1319 (2016).

\bibitem{takusagawa2017projection}
H.~L. Takusagawa, L.~Liu, K.~N. Ma, Y.~Jia, S.~S. Gao, M.~Zhang, B.~Edmunds,
  M.~Parikh, S.~Tehrani, J.~C. Morrison \emph{et~al.},
  \enquote{Projection-resolved optical coherence tomography angiography of
  macular retinal circulation in glaucoma,}
  {\protect\JournalTitle{Ophthalmology}} \textbf{124}, 1589--1599 (2017).

\bibitem{rao2016regional}
H.~L. Rao, Z.~S. Pradhan, R.~N. Weinreb, H.~B. Reddy, M.~Riyazuddin, S.~Dasari,
  M.~Palakurthy, N.~K. Puttaiah, D.~A. Rao, and C.~A. Webers, \enquote{Regional
  comparisons of optical coherence tomography angiography vessel density in
  primary open-angle glaucoma,} {\protect\JournalTitle{American journal of
  ophthalmology}} \textbf{171}, 75--83 (2016).

\bibitem{patel2018plexus}
R.~C. Patel, J.~Wang, T.~S. Hwang, M.~Zhang, S.~S. Gao, M.~E. Pennesi, S.~T.
  Bailey, B.~J. Lujan, X.~Wang, D.~J. Wilson \emph{et~al.},
  \enquote{Plexus-specific detection of retinal vascular pathologic conditions
  with projection-resolved oct angiography,}
  {\protect\JournalTitle{Ophthalmology Retina}} \textbf{2}, 816--826 (2018).

\bibitem{tsuboi2019collateral}
K.~Tsuboi, H.~Sasajima, and M.~Kamei, \enquote{Collateral vessels in branch
  retinal vein occlusion: anatomic and functional analyses by oct angiography,}
  {\protect\JournalTitle{Ophthalmology Retina}} \textbf{3}, 767--776 (2019).

\bibitem{de2015review}
T.~E. De~Carlo, A.~Romano, N.~K. Waheed, and J.~S. Duker, \enquote{A review of
  optical coherence tomography angiography (octa),}
  {\protect\JournalTitle{International journal of retina and vitreous}}
  \textbf{1}, 5 (2015).

\bibitem{jia2017wide}
Y.~Jia, J.~M. Simonett, J.~Wang, X.~Hua, L.~Liu, T.~S. Hwang, and D.~Huang,
  \enquote{Wide-field oct angiography investigation of the relationship between
  radial peripapillary capillary plexus density and nerve fiber layer
  thickness,} {\protect\JournalTitle{Investigative ophthalmology \& visual
  science}} \textbf{58}, 5188--5194 (2017).

\bibitem{ishibazawa2019retinal}
A.~Ishibazawa, L.~R. De~Pretto, A.~Y. Alibhai, E.~M. Moult, M.~Arya, O.~Sorour,
  N.~Mehta, C.~R. Baumal, A.~J. Witkin, A.~Yoshida \emph{et~al.},
  \enquote{Retinal nonperfusion relationship to arteries or veins observed on
  widefield optical coherence tomography angiography in diabetic retinopathy,}
  {\protect\JournalTitle{Investigative Ophthalmology \& Visual Science}}
  \textbf{60}, 4310--4318 (2019).

\bibitem{you2019detection}
Q.~S. You, Y.~Guo, J.~Wang, X.~Wei, A.~Camino, P.~Zang, C.~J. Flaxel, S.~T.
  Bailey, D.~Huang, Y.~Jia \emph{et~al.}, \enquote{Detection of clinically
  unsuspected retinal neovascularization with wide-field optical cohenrence
  tomography angiography.} {\protect\JournalTitle{Retina (Philadelphia, Pa.)}}
  (2019).

\bibitem{camino2017regression}
A.~Camino, Y.~Jia, G.~Liu, J.~Wang, and D.~Huang, \enquote{Regression-based
  algorithm for bulk motion subtraction in optical coherence tomography
  angiography,} {\protect\JournalTitle{Biomedical optics express}} \textbf{8},
  3053--3066 (2017).

\bibitem{uji2018multiple}
A.~Uji, S.~Balasubramanian, J.~Lei, E.~Baghdasaryan, M.~Al-Sheikh, E.~Borrelli,
  and S.~R. Sadda, \enquote{Multiple enface image averaging for enhanced
  optical coherence tomography angiography imaging,}
  {\protect\JournalTitle{Acta ophthalmologica}} \textbf{96}, e820--e827 (2018).

\bibitem{camino2016automated}
A.~Camino, M.~Zhang, C.~Dongye, A.~D. Pechauer, T.~S. Hwang, S.~T. Bailey,
  B.~Lujan, D.~J. Wilson, D.~Huang, and Y.~Jia, \enquote{Automated registration
  and enhanced processing of clinical optical coherence tomography
  angiography,} {\protect\JournalTitle{Quantitative imaging in medicine and
  surgery}} \textbf{6}, 391 (2016).

\bibitem{tan2018enhancement}
B.~Tan, A.~Wong, and K.~Bizheva, \enquote{Enhancement of morphological and
  vascular features in oct images using a modified bayesian residual
  transform,} {\protect\JournalTitle{Biomedical optics express}} \textbf{9},
  2394--2406 (2018).

\bibitem{chlebiej2019quality}
M.~Chlebiej, I.~Gorczynska, A.~Rutkowski, J.~Kluczewski, T.~Grzona,
  E.~Pijewska, B.~L. Sikorski, A.~Szkulmowska, and M.~Szkulmowski,
  \enquote{Quality improvement of oct angiograms with elliptical directional
  filtering,} {\protect\JournalTitle{Biomedical optics express}} \textbf{10},
  1013--1031 (2019).

\bibitem{prentavsic2016segmentation}
P.~Prenta{\v{s}}i{\'c}, M.~Heisler, Z.~Mammo, S.~Lee, A.~Merkur, E.~Navajas,
  M.~F. Beg, M.~{\v{S}}arunic, and S.~Lon{\v{c}}ari{\'c}, \enquote{Segmentation
  of the foveal microvasculature using deep learning networks,}
  {\protect\JournalTitle{Journal of biomedical optics}} \textbf{21}, 075008
  (2016).

\bibitem{guo2018mednet}
Y.~Guo, A.~Camino, J.~Wang, D.~Huang, T.~S. Hwang, and Y.~Jia, \enquote{Mednet,
  a neural network for automated detection of avascular area in oct
  angiography,} {\protect\JournalTitle{Biomedical optics express}} \textbf{9},
  5147--5158 (2018).

\bibitem{nagasato2019automated}
D.~Nagasato, H.~Tabuchi, H.~Masumoto, H.~Enno, N.~Ishitobi, M.~Kameoka,
  M.~Niki, and Y.~Mitamura, \enquote{Automated detection of a nonperfusion area
  caused by retinal vein occlusion in optical coherence tomography angiography
  images using deep learning,} {\protect\JournalTitle{PloS one}} \textbf{14}
  (2019).

\bibitem{guo2019automatic}
M.~Guo, M.~Zhao, A.~M. Cheong, H.~Dai, A.~K. Lam, and Y.~Zhou,
  \enquote{Automatic quantification of superficial foveal avascular zone in
  optical coherence tomography angiography implemented with deep learning,}
  {\protect\JournalTitle{Visual Computing for Industry, Biomedicine, and Art}}
  \textbf{2}, 1--9 (2019).

\bibitem{guo2019development}
Y.~Guo, T.~T. Hormel, H.~Xiong, B.~Wang, A.~Camino, J.~Wang, D.~Huang, T.~S.
  Hwang, and Y.~Jia, \enquote{Development and validation of a deep learning
  algorithm for distinguishing the nonperfusion area from signal reduction
  artifacts on oct angiography,} {\protect\JournalTitle{Biomedical optics
  express}} \textbf{10}, 3257--3268 (2019).

\bibitem{lauermann2019automated}
J.~Lauermann, M.~Treder, M.~Alnawaiseh, C.~Clemens, N.~Eter, and F.~Alten,
  \enquote{Automated oct angiography image quality assessment using a deep
  learning algorithm,} {\protect\JournalTitle{Graefe's Archive for Clinical and
  Experimental Ophthalmology}} \textbf{257}, 1641--1648 (2019).

\bibitem{wang2020automated}
J.~Wang, T.~T. Hormel, L.~Gao, P.~Zang, Y.~Guo, X.~Wang, S.~T. Bailey, and
  Y.~Jia, \enquote{Automated diagnosis and segmentation of choroidal
  neovascularization in oct angiography using deep learning,}
  {\protect\JournalTitle{Biomedical Optics Express}} \textbf{11}, 927--944
  (2020).

\bibitem{wang2020robust}
J.~Wang, T.~T. Hormel, Q.~You, Y.~Guo, X.~Wang, L.~Chen, T.~S. Hwang, and
  Y.~Jia, \enquote{Robust non-perfusion area detection in three retinal
  plexuses using convolutional neural network in oct angiography,}
  {\protect\JournalTitle{Biomedical Optics Express}} \textbf{11}, 330--345
  (2020).

\bibitem{kim2016accurate}
J.~Kim, J.~Kwon~Lee, and K.~Mu~Lee, \enquote{Accurate image super-resolution
  using very deep convolutional networks,} in \emph{Proceedings of the IEEE
  conference on computer vision and pattern recognition,}  (2016), pp.
  1646--1654.

\bibitem{ledig2017photo}
C.~Ledig, L.~Theis, F.~Husz{\'a}r, J.~Caballero, A.~Cunningham, A.~Acosta,
  A.~Aitken, A.~Tejani, J.~Totz, Z.~Wang \emph{et~al.},
  \enquote{Photo-realistic single image super-resolution using a generative
  adversarial network,} in \emph{Proceedings of the IEEE conference on computer
  vision and pattern recognition,}  (2017), pp. 4681--4690.

\bibitem{tong2017image}
T.~Tong, G.~Li, X.~Liu, and Q.~Gao, \enquote{Image super-resolution using dense
  skip connections,} in \emph{Proceedings of the IEEE International Conference
  on Computer Vision,}  (2017), pp. 4799--4807.

\bibitem{xu2018dense}
J.~Xu, Y.~Chae, B.~Stenger, and A.~Datta, \enquote{Dense bynet: Residual dense
  network for image super resolution,} in \emph{2018 25th IEEE International
  Conference on Image Processing (ICIP),}  (IEEE, 2018), pp. 71--75.

\bibitem{zhang2019deep}
K.~Zhang, W.~Zuo, and L.~Zhang, \enquote{Deep plug-and-play super-resolution
  for arbitrary blur kernels,} in \emph{Proceedings of the IEEE Conference on
  Computer Vision and Pattern Recognition,}  (2019), pp. 1671--1681.

\bibitem{jia2012split}
Y.~Jia, O.~Tan, J.~Tokayer, B.~Potsaid, Y.~Wang, J.~J. Liu, M.~F. Kraus,
  H.~Subhash, J.~G. Fujimoto, J.~Hornegger \emph{et~al.},
  \enquote{Split-spectrum amplitude-decorrelation angiography with optical
  coherence tomography,} {\protect\JournalTitle{Optics express}} \textbf{20},
  4710--4725 (2012).

\bibitem{guo2018automated}
Y.~Guo, A.~Camino, M.~Zhang, J.~Wang, D.~Huang, T.~Hwang, and Y.~Jia,
  \enquote{Automated segmentation of retinal layer boundaries and capillary
  plexuses in wide-field optical coherence tomographic angiography,}
  {\protect\JournalTitle{Biomedical optics express}} \textbf{9}, 4429--4442
  (2018).

\bibitem{nair2010rectified}
V.~Nair and G.~E. Hinton, \enquote{Rectified linear units improve restricted
  boltzmann machines,} in \emph{Proceedings of the 27th international
  conference on machine learning (ICML-10),}  (2010), pp. 807--814.

\bibitem{klein2009elastix}
S.~Klein, M.~Staring, K.~Murphy, M.~A. Viergever, and J.~P. Pluim,
  \enquote{Elastix: a toolbox for intensity-based medical image registration,}
  {\protect\JournalTitle{IEEE transactions on medical imaging}} \textbf{29},
  196--205 (2009).

\bibitem{hore2010image}
A.~Hore and D.~Ziou, \enquote{Image quality metrics: Psnr vs. ssim,} in
  \emph{2010 20th International Conference on Pattern Recognition,}  (IEEE,
  2010), pp. 2366--2369.

\bibitem{wang2004image}
Z.~Wang, A.~C. Bovik, H.~R. Sheikh, and E.~P. Simoncelli, \enquote{Image
  quality assessment: from error visibility to structural similarity,}
  {\protect\JournalTitle{IEEE transactions on image processing}} \textbf{13},
  600--612 (2004).

\bibitem{kingma2014adam}
D.~P. Kingma and J.~Ba, \enquote{Adam: A method for stochastic optimization,}
  {\protect\JournalTitle{arXiv preprint arXiv:1412.6980}}  (2014).

\bibitem{peli1990contrast}
E.~Peli, \enquote{Contrast in complex images,} {\protect\JournalTitle{JOSA A}}
  \textbf{7}, 2032--2040 (1990).

\bibitem{otsu1979threshold}
N.~Otsu, \enquote{A threshold selection method from gray-level histograms,}
  {\protect\JournalTitle{IEEE transactions on systems, man, and cybernetics}}
  \textbf{9}, 62--66 (1979).

\bibitem{ogurtsova2017idf}
K.~Ogurtsova, J.~da~Rocha~Fernandes, Y.~Huang, U.~Linnenkamp, L.~Guariguata,
  N.~H. Cho, D.~Cavan, J.~Shaw, and L.~Makaroff, \enquote{Idf diabetes atlas:
  Global estimates for the prevalence of diabetes for 2015 and 2040,}
  {\protect\JournalTitle{Diabetes research and clinical practice}}
  \textbf{128}, 40--50 (2017).

\bibitem{mo2017visualization}
S.~Mo, E.~Phillips, B.~D. Krawitz, R.~Garg, S.~Salim, L.~S. Geyman,
  E.~Efstathiadis, J.~Carroll, R.~B. Rosen, and T.~Y. Chui,
  \enquote{Visualization of radial peripapillary capillaries using optical
  coherence tomography angiography: the effect of image averaging,}
  {\protect\JournalTitle{PloS one}} \textbf{12} (2017).

\bibitem{maloca2017enhanced}
P.~M. Maloca, R.~F. Spaide, S.~Rothenbuehler, H.~P. Scholl, T.~Heeren, J.~E.
  Ramos~de Carvalho, M.~Okada, P.~W. Hasler, C.~Egan, and A.~Tufail,
  \enquote{Enhanced resolution and speckle-free three-dimensional printing of
  macular optical coherence tomography angiography,}
  {\protect\JournalTitle{Acta Ophthalmol}} pp. 1--3 (2017).

\bibitem{hendargo2013automated}
H.~C. Hendargo, R.~Estrada, S.~J. Chiu, C.~Tomasi, S.~Farsiu, and J.~A. Izatt,
  \enquote{Automated non-rigid registration and mosaicing for robust imaging of
  distinct retinal capillary beds using speckle variance optical coherence
  tomography,} {\protect\JournalTitle{Biomedical optics express}} \textbf{4},
  803--821 (2013).

\bibitem{kashani2017optical}
A.~H. Kashani, C.-L. Chen, J.~K. Gahm, F.~Zheng, G.~M. Richter, P.~J.
  Rosenfeld, Y.~Shi, and R.~K. Wang, \enquote{Optical coherence tomography
  angiography: A comprehensive review of current methods and clinical
  applications,} {\protect\JournalTitle{Progress in retinal and eye research}}
  \textbf{60}, 66--100 (2017).

\bibitem{spaide2018optical}
R.~F. Spaide, J.~G. Fujimoto, N.~K. Waheed, S.~R. Sadda, and G.~Staurenghi,
  \enquote{Optical coherence tomography angiography,}
  {\protect\JournalTitle{Progress in retinal and eye research}} \textbf{64},
  1--55 (2018).

\bibitem{russell2019distribution}
J.~F. Russell, H.~W. Flynn~Jr, J.~Sridhar, J.~H. Townsend, Y.~Shi, K.~C. Fan,
  N.~L. Scott, J.~W. Hinkle, C.~Lyu, G.~Gregori \emph{et~al.},
  \enquote{Distribution of diabetic neovascularization on ultra-widefield
  fluorescein angiography and on simulated widefield oct angiography,}
  {\protect\JournalTitle{American journal of ophthalmology}} \textbf{207},
  110--120 (2019).

\bibitem{li2017adaptive}
P.~Li, Z.~Huang, S.~Yang, X.~Liu, Q.~Ren, and P.~Li, \enquote{Adaptive
  classifier allows enhanced flow contrast in oct angiography using a
  histogram-based motion threshold and 3d hessian analysis-based shape
  filtering,} {\protect\JournalTitle{Optics letters}} \textbf{42}, 4816--4819
  (2017).

\bibitem{ting2019artificial}
D.~S.~W. Ting, L.~R. Pasquale, L.~Peng, J.~P. Campbell, A.~Y. Lee, R.~Raman,
  G.~S.~W. Tan, L.~Schmetterer, P.~A. Keane, and T.~Y. Wong,
  \enquote{Artificial intelligence and deep learning in ophthalmology,}
  {\protect\JournalTitle{British Journal of Ophthalmology}} \textbf{103},
  167--175 (2019).

\bibitem{zeiler2014visualizing}
M.~D. Zeiler and R.~Fergus, \enquote{Visualizing and understanding
  convolutional networks,} in \emph{European conference on computer vision,}
  (Springer, 2014), pp. 818--833.

\bibitem{zhang2018interpretable}
Q.~Zhang, Y.~Nian~Wu, and S.-C. Zhu, \enquote{Interpretable convolutional
  neural networks,} in \emph{Proceedings of the IEEE Conference on Computer
  Vision and Pattern Recognition,}  (2018), pp. 8827--8836.

\bibitem{samek2017explainable}
W.~Samek, T.~Wiegand, and K.-R. M{\"u}ller, \enquote{Explainable artificial
  intelligence: Understanding, visualizing and interpreting deep learning
  models,} {\protect\JournalTitle{arXiv preprint arXiv:1708.08296}}  (2017).

\bibitem{zhang2018visual}
Q.-s. Zhang and S.-C. Zhu, \enquote{Visual interpretability for deep learning:
  a survey,} {\protect\JournalTitle{Frontiers of Information Technology \&
  Electronic Engineering}} \textbf{19}, 27--39 (2018).

\bibitem{zintgraf2017visualizing}
L.~M. Zintgraf, T.~S. Cohen, T.~Adel, and M.~Welling, \enquote{Visualizing deep
  neural network decisions: Prediction difference analysis,}
  {\protect\JournalTitle{arXiv preprint arXiv:1702.04595}}  (2017).

\bibitem{toto2016retinal}
L.~Toto, E.~Borrelli, L.~Di~Antonio, P.~Carpineto, and R.~Mastropasqua,
  \enquote{Retinal vascular plexuses'changes in dry age-related macular
  degeneration, evaluated by means of optical coherence tomography
  angiography,} {\protect\JournalTitle{Retina}} \textbf{36}, 1566--1572 (2016).

\bibitem{chi2017optical}
Y.-T. Chi, C.-H. Yang, and C.-K. Cheng, \enquote{Optical coherence tomography
  angiography for assessment of the 3-dimensional structures of polypoidal
  choroidal vasculopathy,} {\protect\JournalTitle{JAMA ophthalmology}}
  \textbf{135}, 1310--1316 (2017).

\bibitem{onishi2018importance}
A.~C. Onishi, P.~L. Nesper, P.~K. Roberts, G.~A. Moharram, H.~Chai, L.~Liu,
  L.~M. Jampol, and A.~A. Fawzi, \enquote{Importance of considering the middle
  capillary plexus on oct angiography in diabetic retinopathy,}
  {\protect\JournalTitle{Investigative ophthalmology \& visual science}}
  \textbf{59}, 2167--2176 (2018).

\bibitem{hwang2018automated}
T.~S. Hwang, A.~M. Hagag, J.~Wang, M.~Zhang, A.~Smith, D.~J. Wilson, D.~Huang,
  and Y.~Jia, \enquote{Automated quantification of nonperfusion areas in 3
  vascular plexuses with optical coherence tomography angiography in eyes of
  patients with diabetes,} {\protect\JournalTitle{JAMA ophthalmology}}
  \textbf{136}, 929--936 (2018).

\bibitem{camino2019detecting}
A.~Camino, Y.~Guo, Q.~You, J.~Wang, D.~Huang, S.~T. Bailey, and Y.~Jia,
  \enquote{Detecting and measuring areas of choriocapillaris low perfusion in
  intermediate, non-neovascular age-related macular degeneration,}
  {\protect\JournalTitle{Neurophotonics}} \textbf{6}, 041108 (2019).

\bibitem{liu2019projection}
L.~Liu, B.~Edmunds, H.~L. Takusagawa, S.~Tehrani, L.~H. Lombardi, J.~C.
  Morrison, Y.~Jia, and D.~Huang, \enquote{Projection-resolved optical
  coherence tomography angiography of the peripapillary retina in glaucoma,}
  {\protect\JournalTitle{American journal of ophthalmology}} \textbf{207},
  99--109 (2019).

\bibitem{zang2016automated}
P.~Zang, G.~Liu, M.~Zhang, C.~Dongye, J.~Wang, A.~D. Pechauer, T.~S. Hwang,
  D.~J. Wilson, D.~Huang, D.~Li \emph{et~al.}, \enquote{Automated motion
  correction using parallel-strip registration for wide-field en face oct
  angiogram,} {\protect\JournalTitle{Biomedical optics express}} \textbf{7},
  2823--2836 (2016).

\end{thebibliography}
	
\end{document}